\documentclass[twocolumn]{aastex62}
\usepackage{natbib}
\usepackage{amsmath}
\usepackage{textcomp}
\usepackage{booktabs}
\usepackage{bookmark}
\usepackage{color}
\usepackage{soul}
\usepackage{url}
\usepackage{enumitem}
\usepackage{CJK}
\usepackage{mathrsfs}  
\usepackage[version=4]{mhchem}

\hypersetup{linkcolor=blue, citecolor=blue}
\shorttitle{Bivariate Luminosity Function of Galaxy Pairs}
\shortauthors{Feng ET AL.}

\begin{document}
\begin{CJK*}{UTF8}{gbsn}
\newcommand{\kpc}{\ h^{-1}\ \textmd{kpc}}
\newcommand{\kms}{\ \textmd{km}\ \textrm{s}^{-1}}
\newcommand{\pdis}{d_{\textrm{p}}}

\title{Bivariate Luminosity Function of Galaxy Pairs}

\author[0000-0002-9767-9237]{Shuai Feng (冯帅)}
\affiliation{Key Laboratory for Research in Galaxies and Cosmology, Shanghai Astronomical Observatory, Chinese Academy of Sciences, \\
80 Nandan Road, Shanghai 200030, China}
\affiliation{University of the Chinese Academy of Sciences, No.19A Yuquan Road, Beijing 100049, China}

\author{Shi-Yin Shen (沈世银)}
\affiliation{Key Laboratory for Research in Galaxies and Cosmology, Shanghai Astronomical Observatory, Chinese Academy of Sciences, \\
80 Nandan Road, Shanghai 200030, China}
\affiliation{Key Lab for Astrophysics, Shanghai 200234, China}

\author{Fang-Ting Yuan (袁方婷)}
\affiliation{Key Laboratory for Research in Galaxies and Cosmology, Shanghai Astronomical Observatory, Chinese Academy of Sciences, \\
80 Nandan Road, Shanghai 200030, China}

\author{A-Li Luo (罗阿里)}
\author{Jian-Nan Zhang (张健楠)}
\author{Meng-Xin Wang (汪梦欣)}
\author{Xia Wang (汪霞)}
\author{Yin-Bi Li (李荫碧)}
\author{Wen Hou (侯文)}
\author{Xiao Kong (孔啸)}
\author{Yan-Xin Guo (郭炎鑫)}
\author{Fang Zuo (左芳)}
\affiliation{Key Laboratory of Optical Astronomy, National Astronomical Observatories, Chinese Academy of Sciences, Beijing 100101, China}

\correspondingauthor{Shi-Yin Shen}
\email{ssy@shao.ac.cn}

\begin{abstract}
We measure the bivariate luminosity function (BLF) of galaxy pairs and use it to probe and characterize the galaxy-galaxy interaction between pair members. The galaxy pair sample is selected from the main galaxy sample of Sloan Digital Sky Survey and supplied with a significant number of redshifts from the LAMOST spectral and GAMA surveys. We find the BLFs depend on the projected distance $\pdis$ between pair members. At large separation $\pdis > 150 \kpc$, the BLF degenerates into a luminosity function (LF) of single galaxies, indicating few interactions between pair members. At $100 \kpc \leq \pdis \leq 150 \kpc$, the BLF starts to show the correlation between pair members, in the sense that the shape of the conditional luminosity function (CLF) of one member galaxy starts to depend on the luminosity of the other member galaxy. Specifically, the CLF with a brighter companion has a steeper faint-end slope, which becomes even more significant at $50 \kpc \leq \pdis \leq 100 \kpc$. This behavior is consistent with the scenario, \textit{and also is the  observational evidence}, that dynamic friction drives massive major merger pairs to merge more quickly. At close distance $\pdis \leq 50 \kpc$, besides the merging time-scale effect, the BLF also shows an overall brightening of $\Delta M_r \geq 0.04$ mag, which reveals the enhanced star formation of the close-pair phase. By combining another statistical conclusion that the star formation rate of late-type galaxies in close pairs is enhanced at a level of about 40\%, we further conclude that the average starburst time-scale of close pairs is as long as 0.4 Gyr.
\end{abstract}

\keywords{galaxies: interaction -- galaxies: luminosity function, mass function}

% =========================

\section{introduction}\label{sec1}

In a hierarchical galaxy formation diagram, the galaxy-galaxy interaction plays a key role in shaping the galaxy evolution path. As the simplest galaxy `systems', galaxy pairs are the ideal places to study the interactions between galaxies. Recent decades have proven that the interaction in close pairs (e.g., $\pdis \leq 50 \kpc$) could change the physical properties of galaxies significantly. Compared to field galaxies, the galaxies with close encounters typically have disturbed morphologies \citep{toomre1972,hernandez-toledo2005,hernandez2006,Patton2016}, enhanced star formation rates \citep{barton2000,woods2006,ellison2008,scudder2012,Davies2015}, diluted nuclear metallicities \citep{ellison2008,kewley2010,scudder2012}, and overabundances of active galactic nuclei \citep{ellison2008,ellison2011,liu2011,fu2018}. On the other hand, it is far from clear how the interaction in wide pairs (e.g., $\pdis \geq 50 \kpc$) influences galaxy evolution. Those wide pairs do not show significant differences from the field galaxies, such as morphological asymmetry and enhanced star formation rate \citep{patton2013,Patton2016}.

Besides changing their physical properties, galaxy-galaxy interaction also shapes galaxy evolution through galaxy merging. In galaxy pairs, the orbital energy and angular momentum of pair members gradually dissipate through dynamic friction and then finally merge. Numerical simulations show, that the typical merging time-scale is about $1-2$ Gyr and mainly depends on the masses of both pair members\citep{lacey1994,Boylan-Kolchin2000,jiang2008,lotz2010b}. As a result, the relative number density of galaxy pairs would deviate from a random combination of field galaxies, even when the changes in their physical properties seem negligible.

In general, in the galaxy pair environment, the enhanced star formation induced by close interaction (e.g., tidal effect) will brighten the galaxies. On the other hand, if the merging time-scale of more massive galaxies is shorter(e.g., \citealt{Jiang2014}), there would be a lower fraction of bright galaxies in pairs than that in the field. Therefore, the luminosity function of galaxies in pairs would be different from that of field galaxies. Apart from the general difference, the interaction between pair members further complicates the statistical properties of the galaxies in pairs. For example, the galaxies in major merger pairs (those with mass ratio larger than 1:3) are believed to suffer stronger interaction (therefore more enhanced star formation) than in minor ones (e.g., \citealt{Davies2015}). Also, the merging time-scale is shown to be strongly dependent on the mass ratio of pair members(e.g., \citealt{Boylan-Kolchin2000}). Therefore, as a result of these complex interactions, the physical and statistical properties of pair members would not only deviate from those of single galaxies but would also be strongly coupled to each other.

The luminosity function (LF), as a basic statistical description of the relative number of galaxies with different luminosity, provides a fundamental constraint on the modeling of galaxy formation and evolution. For galaxy pairs, there are two different types of LFs, the univariate LF(ULF) and bivariate LF(BLF). The ULF does not distinguish the pair members and provides a measurement of the pair fraction of galaxies \citep{xu1991,xu2004,domingue2009}. For example, \citet{domingue2009} found that the ULF of galaxy pairs in $K_S$ band is very similar to the global LF of the galaxy sample, which implies a constant pair fraction of $\sim 1.6\%$. Comparing with ULF, BLF contains more information, in particular, on the correlations of variables\citep{Sodre1993,ball2006,takeuchi2013}. \citet{Robotham2014} investigated the fraction of galaxy pairs with different mass configurations using a bivariate measurement of stellar mass distribution of pair members. However, the internal correlation between pair members has not been addressed.

To have a comprehensive description of galaxy pairs, where both the configuration of galaxy pairs(e.g., merging time-scale) and the physical properties of pair members(e.g., enhanced star formation) is taken into account, the BLF of pair members $\Phi(M_A,M_B)$ might be an ideal diagnostic tool. If we assume that the galaxy pairs are initially built from random pairing of single galaxies, we would expect that the initial BLF of galaxy pairs could degenerate into the LF of single galaxies, in the sense that the shape of the conditional luminosity function (CLF) of one member galaxy $\Phi(M_A|M_B)$ would not depend on its companion luminosity $M_B$. Therefore, this assumption could be used to test the critical distance where the galaxy-galaxy interaction starts. As the galaxy pairs evolve, the physical interactions change the luminosities of pair members, whereas the merging process further alters their relative number densities. Both mechanisms will change and shape the BLF of galaxy pairs. 

To have an accurate measurement of $\Phi(M_A,M_B)$, a large and well-defined galaxy pair sample is required. The largest galaxy pair sample at low redshift to date is built based on the main galaxy sample of the Sloan Digital Sky Survey (SDSS, \citealt{york2000}), e.g., \citet{ellison2008}. In this study, we will also build the galaxy pair sample based on SDSS galaxies. Compared with the previous works, we have two main improvements. First, we have supplied a large number of redshifts from the LAMOST spectral survey\citep{Shen2016}, which increases both the total number and completeness of the galaxy pair sample(Section \ref{sec_data}). Second, we extend the projected distance threshold out to 200$\kpc$ during galaxy pair selection. This threshold is more extended than most of the other studies ($\sim 100\kpc$, e.g.,\citealt{scudder2012}), where the galaxy-galaxy interaction effects are only found at smaller separation. 

This paper is organized as follows. In Section \ref{sec_data} we introduce the galaxy sample, criteria of galaxy pair selection and incompleteness correction. Then, we measure and parameterize the BLFs of galaxy pairs as a function of projection distance in Section \ref{sec_blf}. In Section \ref{sec_model}, we build toy models to discuss the physical implications of the BLFs. Finally, we give a summary in Section \ref{sec_sum}. Throughout this paper, we assume standard $\Lambda$-CDM concordance cosmology with $\Omega_m=0.3, \Omega_{\Lambda}=0.7$ and $h = H_0\ /\ 100 \kms\ \textmd{Mpc}^{-1}$.

Throughout this paper, we use $L_A$ and $L_B$ to denote luminosities of pair members, where $A$ and $B$ are assigned to the pair members randomly. On the other hand, when we consider luminosities of the primary and secondary galaxies, we use $L_1$ and $L_2$  instead. We denote the mass of pair members in a similar way.

\section{data}\label{sec_data}

In this section, we describe the parent galaxy sample, the selection algorithm for galaxy pairs and the correction for spectroscopic incompleteness.

\subsection{Parent Sample}\label{sect_parent_sample}

The parent galaxy catalog we use is the New York University Value-Added Galaxy Catalog (NYU-VAGC; \citealt{blanton2005}) of the SDSS DR7\citep{Abazajian2008}, which is a cross-matched collection of galaxy catalogs including extra redshifts from other surveys(e.g., 2dF, PSCz, RC3). To have high redshift completeness, following the main galaxy sample(MGS, \citealt{strauss2002}) of SDSS, we take the galaxies with Galactic extinction corrected magnitude in the range of $14.0\leq m_r \leq 17.77$. In NYU-VAGC, the number of galaxies in this magnitude range is $746,950$. We refer to these galaxies as the parent photometric sample. Among them, $696,245$ have redshifts, which is contributed by different surveys, but mainly from the SDSS DR7($686,398$, $>98.6\%$). The small fraction of galaxies without redshifts in the SDSS MGS is primarily caused by the fiber collision effect \citep{Shen2016}.

The MGS galaxies without redshifts in the SDSS DR7 were continually targeted in the SDSS III and IV with spare fibers. We matched the parent photometric sample with the latest SDSS data release, DR14 \citep{SDSSDR14}\footnote{There are no new extra redshifts in the SDSS DR15.} and obtained $8,532$ extra redshifts. Moreover, these MGS galaxies without redshifts in the SDSS DR7 have also been compiled as a complementary galaxy sample and targeted in the LAMOST spectral survey(\citealt{cui2012,zhao2012,luo2015}). In the LAMOST data release 5, we find an additional $7,004$ galaxies with spectroscopic redshifts. We also matched the parent photometric sample with data release 2 of the GAMA survey \citep{gamadr2} and obtained a further 1,024 redshifts. For a few galaxies having more than one redshifts measured by different surveys, we set the priority as follows: SDSS $>$ LAMOST DR5 $>$ GAMA DR2 $>$  others in NYU-VAGC. Finally, we obtained redshifts for $711,347$ galaxies in total, which results in spectroscopic completeness of $95.2\%$. The exact numbers of redshifts obtained from different surveys are listed in Table \ref{tb1}.

\begin{table}[ht]
\centering
\caption{Redshifts of the parent sample.}
\label{tb1}
\begin{tabular}{lr}
\hline
 Redshift Catalog & Redshift Number     \\
\hline
 SDSS DR14                    & 694,930    \\
 LAMOST DR5                 & 7,004        \\
 GAMA DR2                   & 1,024      \\
 Others from NYU-VAGC        & 8,339        \\
\hline
 Spectroscopic Sample in MGS    & 711,347    \\
 Photometric Sample in MGS        & 746,950    \\
\hline
\end{tabular}
\end{table}

\subsection{Pair Selection}\label{sec_pairselection}

\begin{figure*}[htp]
  \centering
  \includegraphics[width=180mm]{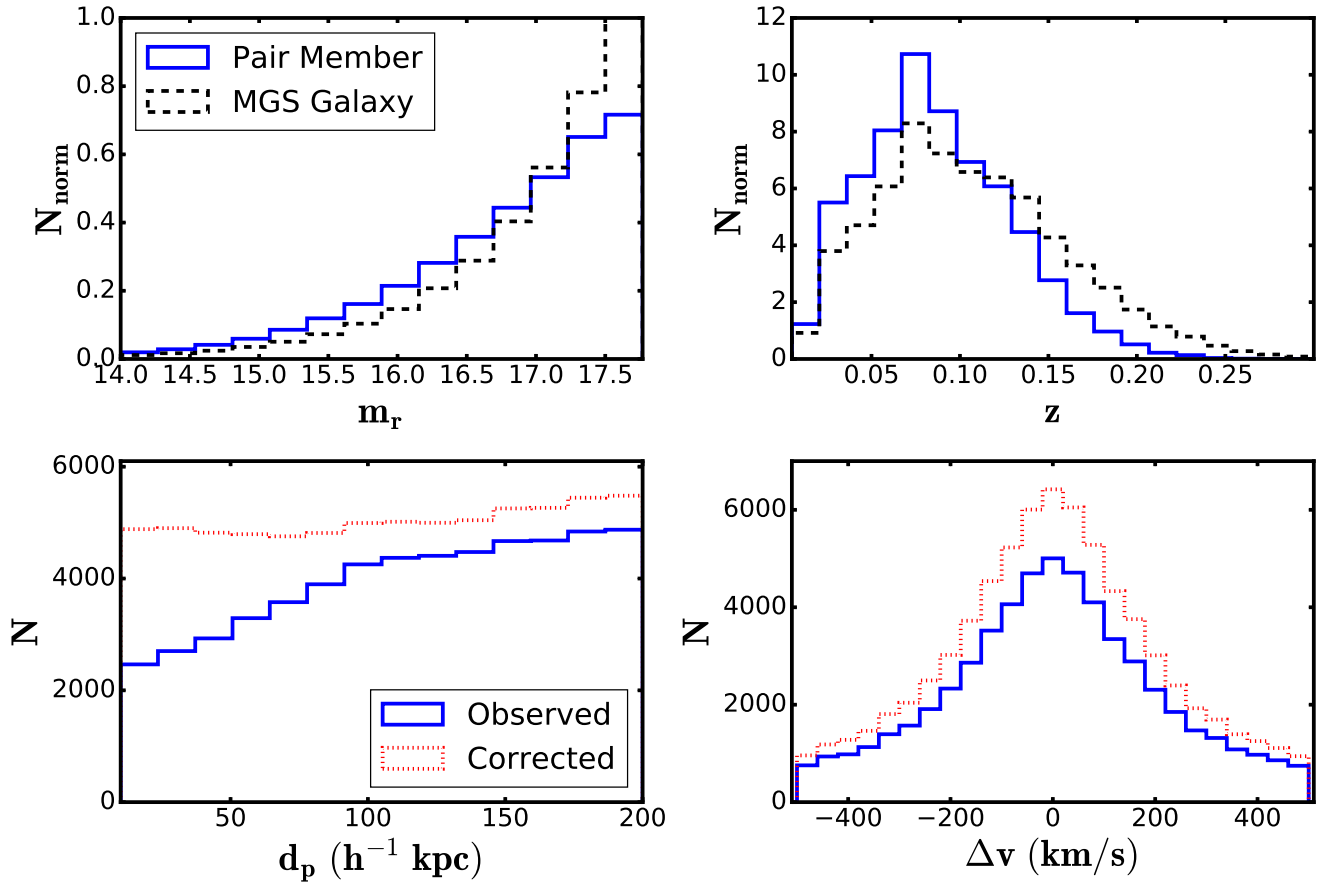}
  \caption{\textit{Top}: the apparent magnitude and redshift distributions of the pair members (blue solid lines) and parent sample (black dashed lines). \textit{Bottom}: the projected distance and velocity difference distributions of the galaxy pair sample (blue solid lines), where the red dotted lines show the corresponding distributions after spectroscopic incompleteness correction. }
  \label{fig_samp}
\end{figure*}

We use a friend-of-friend (FOF) algorithm to select galaxy pairs from the parent sample. The main selection criteria are listed as follows:
\begin{enumerate}
\item The line-of-sight velocity difference of two galaxies satisfies $|\Delta v| \leq 500 \kms$.
\item The projected distance between two galaxies is in the range of $10 \kpc \leq \pdis \leq 200 \kpc$.
\item Each galaxy has only one neighbor satisfying the above criteria.
\end{enumerate}

We exclude the very close galaxy pairs ($\pdis < 10 \kpc$) to avoid the very peculiar morphology and unreliable deblending photometry. We require that each galaxy only has one neighbor to avoid those galaxies in a compact group environment.  Since our galaxy sample is still not 100 \% complete in spectroscopy, there are $9,175$ galaxy pairs having neighbors within $10 \kpc \leq \pdis \leq 200 \kpc$, and these neighbors do not have redshifts. In this case, some of the pairs would violate the `isolated' criteria if their neighbors also have the same redshifts as them. For the sake of safety, we have not considered these galaxy pairs in the BLF calculation.  Finally, we have $46,510$ galaxy pairs.

We also have cross-matched the galaxy pair sample with the SDSS group catalog of \citet{yang2007}. We find that less than one fifth ($9,188/46,510$) of them are located in the groups with more than two members. For these galaxy pairs in \textit{groups}, since the other members are located much further away $\pdis > 200 \kpc$, the influences of external interaction from other members at such distances is negligible (see Section \ref{sec_BLF_pdis}). We also  have tested that the exclusion of these pairs in \textit{groups} would not change any of our conclusions except to increase the uncertainty.

In Figure \ref{fig_samp}, we show the distributions of some basic parameters of the galaxy pair sample, including the $r$-band magnitudes and redshifts of pair members, the projected distances and velocity differences of galaxy pairs. In each of the top two panels, the global distribution of the parent galaxy sample is overlayed as a dashed histogram for comparison. The galaxies in pairs show significant differences from the parent galaxy sample. These differences are mainly caused by the fiber collision effect, where the more distant(fainter) pairs are more likely to be missed because of their closer angular separation. In the bottom two panels, we show $\pdis$ and $\Delta V$ distribution of the galaxy pair sample with solid blue lines. As comparison, we also show the $\pdis$ and $\Delta V$ distribution of galaxy pairs after spectroscopic completeness correction (see Section \ref{sec_spec_completeness}) as red dotted lines. After correction of the incompleteness, the $\pdis$ distribution shows a quite flat distribution while the $\Delta V$ distribution is almost unchanged.

\subsection{Spectroscopic Incompleteness} \label{sec_spec_completeness}

\begin{figure*}
\centering
\includegraphics[width=180mm]{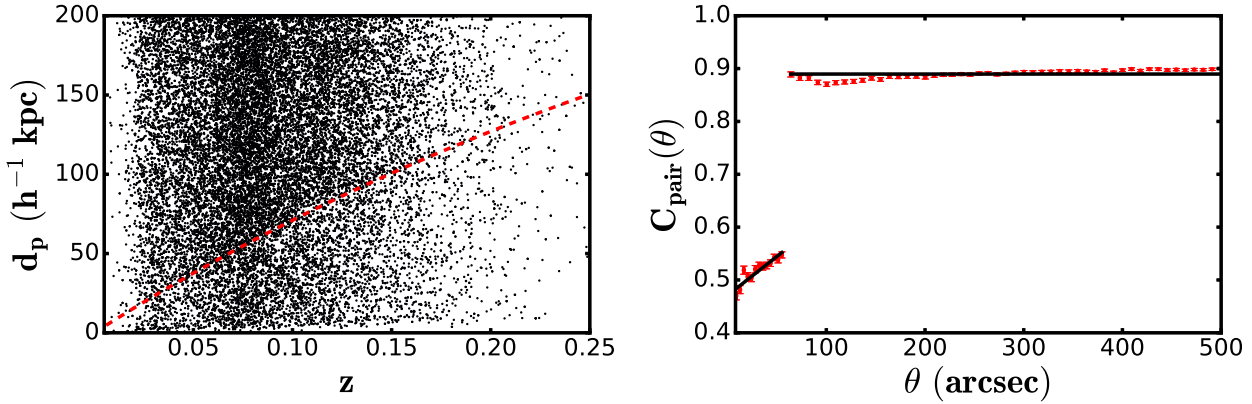}
\caption{\textit{Left panel}: the distribution of the galaxy pair sample in the $\pdis-z$ plane, in which the red dashed line represents $\theta=55{''}$. \textit{Right panel}: the spectroscopic completeness as a function of angular separation $\theta$. The black solid line shows the fitting formula of Equation \ref{eq_Ctheta}. The error bars show the uncertainties of $68\%$ confidence from Possion statistics. }
\label{fig_spec_comp}
\end{figure*}

Although the completeness of the parent galaxy sample has been improved by supplementing lots of additional redshifts (Table \ref{tb1}), spectroscopic incompleteness still remains, especially for very close pairs. To demonstrate this effect, we show the distribution of the galaxy pair sample in $\pdis-z$ plane(left panel of Figure \ref{fig_spec_comp}). The red line in Figure \ref{fig_spec_comp} shows the $\pdis-z$ relation for the angular separation of $\theta=55 {''}$, which is the minimum angular separation between any two SDSS fibers \citep{york2000}. The incompleteness at $\theta<55 {''}$ is still obvious.

To make corrections for the incompleteness of the galaxy pair sample, we calculate the spectroscopic completeness of the galaxy pairs with different angular separation as follows. We explore a series of annuli($\theta \pm \Delta \theta/2$) around each galaxy of the parent galaxy sample.  We set $\Delta\theta$ to be 3${''}$ and $\theta$ to span from 10${''}$ to 500${''}$, which corresponds to physical separations of $4\kpc$ and $200\kpc$ at redshift $0.03$ respectively. For given $\theta$, we first count the number of galaxies enclosed by all annuli as $N_{\text{pp}}(\theta)$, which is equivalent to the number of close-projected galaxies. Among them, we further count the number of its subsample as $N_{\text{ss}}(\theta)$, in which both two close-projected galaxies have spectroscopic redshifts, regardless of whether they are real galaxy pairs. Then, the spectroscopic completeness of our galaxy pair sample for the angular separation of $\theta$ can be expressed as
\begin{equation}
C_{\textrm{pair}}(\theta)=\frac{N_{\text{ss}}(\theta)}{N_{\text{pp}}(\theta)}
\end{equation}

We show $C_{\textrm{pair}}(\theta)$ as the dots in the right panel of Figure \ref{fig_spec_comp}, where the error bars show the uncertainties from Poisson statistics. As can be seen, the change of $C_{\text{pair}}(\theta)$ at $\theta=55{''}$ is dramatic. At large separation $\theta > 55{''}$, $C_{\text{pair}}(\theta)$ is approximately a constant $\sim 0.89$, which is roughly equal to the square of the global spectroscopic completeness of our parent sample($95.2\%$). At $\theta<55{''}$, $C_{\text{pair}}(\theta)$ on average is about 50\% , and systemically decreases with decreasing $\theta$. Although the completeness of our galaxy pair sample is still far from $100\%$, it is much better than in previous studies($\sim 30\%$, e.g. \citealt{ellison2008}) where only SDSS DR7 data are available. Since most of our galaxy pairs have $\theta > 55{''}$, the average spectroscopic completeness of the galaxy pair sample reaches 82\%.

For convenience, we fit $C_{\text{pair}}(\theta)$ with a step function,
\begin{equation}\label{eq_Ctheta}
C_{\textrm{pair}}(\theta)=\left\{ \begin{array}{ll}
               0.889\quad\quad & \theta \geq 55{''}\\
               0.0015\theta+0.471\quad\quad & \theta < 55{''}
              \end{array}
       \right.
\end{equation}
which is also shown as the solid line in the right panel of Figure \ref{fig_spec_comp}.

Besides the average completeness function shown by Equation \ref{eq_Ctheta}, it should be emphasized that the completeness function of our galaxy pair sample varies over the sky. For example, the completeness in the sky coverage of the GAMA survey ($\sim$ 290 deg$^2$) is much higher than other areas since the GAMA has the completeness of $>98\%$ down to $r<17.77$ \citep{baldry2010}. However, considering the extensive sky coverage of our galaxy pair sample ($\sim7000$ deg$^2$), we do not expect any cosmic variance effect would alter the statistical conclusions of this study.

\section{bivariate luminosity function} \label{sec_blf}

In this section, we first introduce the algorithm of the BLF calculation and then present the BLF dependence on the projected distance of pair members. Next, we introduce an analytical formula to parametrize the shape of the measured BLFs. At the end of this section, we discuss the ULF and the number density of galaxy pairs.

To build the LF, we calculate the $K+E$ corrected absolute magnitude of each galaxy at $z=0.1$ following \cite{blanton2003a}, 
\begin{equation*}
M_r^{0.1}=m_r-5\log(d_L/10 \text{pc})-K_r^{0.1}(z)+(z-0.1)Q
\end{equation*}
where $m_r$ is the $r$-band apparent magnitude corrected for Galactic extinction, $d_L$ is the luminosity distance, $K_r(z)$ is K-correction value calculated using the \texttt{KCORRECT} algorithm of \citet{blanton2007}, and $Q$ is the evolution correction which equals to $1.62$ for $r-$band \citep{blanton2003a}.

\subsection{Method}\label{sec_eep2d}

We take the extension of the non-parametric step-wise maximum likelihood estimator(SWML, \citealt{efstathiou1988}) for BLF calculation \citep{Sodre1993}. In this method, the BLF of galaxy pairs $\Phi(A,B)$ is defined as a step function,
\begin{equation}
\Phi(A,B)=\phi_{jk}\qquad(j=1,...,N_A;\ k=1,...,N_B)
\end{equation}
When the magnitudes of pair members $A$ and $B$ locate in absolute magnitude bin $A_j^-<A<A_k^+$ and $B_k^-<B<B_k^+$ respectively, the value of pair BLF equals to $\phi_{jk}$. $N_A$ and $N_B$ are the magnitude bin numbers, and magnitude bin edges are set as
\begin{equation*}
\begin{split}
A_j^{\pm}=A_j \pm \Delta A/2\ , \\ 
B_k^{\pm}=B_k \pm \Delta B/2\,,
\end{split}
\end{equation*}
where $\Delta A$ and $\Delta B$ are the bin widths.

Considering a magnitude-limited sample containing $N_{\text{p}}$ galaxy pairs with the member absolute magnitudes ($A_i$, $B_i$), angular separation $\theta_i$ and redshift $z_i$ respectively ($i=1,...,N_{\text{p}}$), for given $\Phi(A,B)$, the probability of observing $i$th galaxy pair is  
\begin{equation}\label{eq_p_BLF}
p_i = \frac{\Phi(A_i,B_i)C_{\text{pair}}(\theta_i)}
{\int^{A_{\textmd{bright}}(z_i)}_{A_{\textmd{faint}}(z_i)}
\int^{B_{\textmd{bright}}(z_i)}_{B_{\textmd{faint}}(z_i)}\Phi(A,B)C_{\text{pair}}(\theta)dAdB} \,,
\end{equation}
where $A_{\textmd{bright}}(z_i)$ and $A_{\textmd{faint}}(z_i)$ ($B_{\textmd{bright}}(z_i)$ and $B_{\textmd{faint}}(z_i)$) are the brightest and faintest absolute magnitude that the member of $i$th galaxy pair can be observed for the given apparent magnitude limit $m_{\textmd{bright}}$ and $m_{\textmd{faint}}$. For our galaxy pair sample, the magnitude limits are $m_{\textmd{bright}}=14.00$, $m_{\textmd{faint}}=17.77$. Then the global likelihood $\mathscr{L}$ is defined as the products of $p_i$ for all pairs,
\begin{equation}\label{likelihood}
\mathscr{L}=\prod^{N_{\text{p}}} p_i \,.
\end{equation}

To be specific, $\mathscr{L}$ can be rewritten as follows,
\begin{equation}\label{loglike}
\begin{split}
\ln \mathscr{L} =&
\sum^{N_\text{p}}_{i=1} \sum^{N_A}_{j=1} \sum^{N_B}_{k=1} W_{ijk} \ln[\phi_{jk}C_{\text{pair}}(\theta_i)] \\
& -\sum^{N_\text{p}}_{i=1} \ln(\sum^{N_A}_{j=1}\sum^{N_B}_{k=1}H_{ijk}\phi_{jk})+ \textmd{constant} \,.
\end{split}
\end{equation}
Here, we have defined two window functions,
\begin{equation}
W_{ijk}=\left\{ \begin{array}{ll}
               1\quad& \textmd{if} -\Delta A/2 \leq A_i-A_j < \Delta A/2 \\
                  & \textmd{and} -\Delta B/2 \leq B_i-B_k < \Delta B/2\\
               0\quad & \textmd{otherwise}
              \end{array}
       \right.
\end{equation}
and
\begin{equation}
H_{ijk}=\frac{1}{\Delta A \Delta B}\int^{A'}_{A^-}\int^{B'}_{B^-}C_{\text{pair}}(\theta_i)dAdB
\end{equation}
where
\begin{equation*}
\begin{split}
A'=\max[A^-,\min(A^+,A^i_{\text{faint}})]\ , \\
B'=\max[B^-,\min(B^+,B^i_{\text{faint}})]\,.
\end{split}
\end{equation*}

The maximum value of $\mathscr{L}$ is calculated by solving $\partial \mathscr{L}/ \partial \phi_{jk}=0$. In practice, $\phi_{jk}$ is obtained through iteration:
\begin{equation}
\phi_{jk}=\frac{\sum_{i}^{N_{\text{p}}}W_{ijk}}
{\sum_{i}^{N_{\text{p}}} ( H_{ijk} / \sum_{l}^{N_A} \sum_{m}^{N_B} \phi_{lm} H_{ilm})}\,.
\end{equation}

In the above calculation, the galaxies in pairs are assigned as member $A$ and $B$ randomly. The final BLF value of each bin is taken as the average of the corresponding diagonal bin. In this case, the final BLF satisfies $\Phi(A,B) = \Phi(B,A)$.

Similar to SWML, this method only provides the shape of BLF. The normalized density of the pair BLFs need to be estimated independently. For our galaxy pair sample, the number density of the galaxy pairs is estimated by
\begin{equation}
n_{\text{pair}}=\sum_i^{N_\text{p}}\frac{1}{C_{\textrm{pair}}(\theta_i)V_{i,\textrm{max}}}\,,
\end{equation}
where $V_{i,\textrm{max}}$ is the largest volume where $i$th galaxy pair is observable. Meanwhile, from the definition of BLF, the number density of galaxy pairs could also be expressed as
\begin{equation}
n_{\text{pair}}=\int^{A_{\textrm{max}}}_{A_{\textrm{min}}}
\int^{B_{\textrm{max}}}_{B_{\textrm{min}}}\Phi(A,B)\text{d}A\text{d}B \,.
\end{equation}
With the above two equations, the normalization parameter of BLF could finally be estimated.

The error of BLF is derived using the bootstrap algorithm. We make 100 bootstrap galaxy pair samples, and use the standard deviation of the best estimations of these 100 samples as the uncertainty of the BLF.

% ===================================================================================================

\subsection{BLF Dependence on Projected Distance}\label{sec_BLF_pdis}

\begin{figure*}[htp]
\centering
\begin{tabular}{c}
\includegraphics[width=180mm]{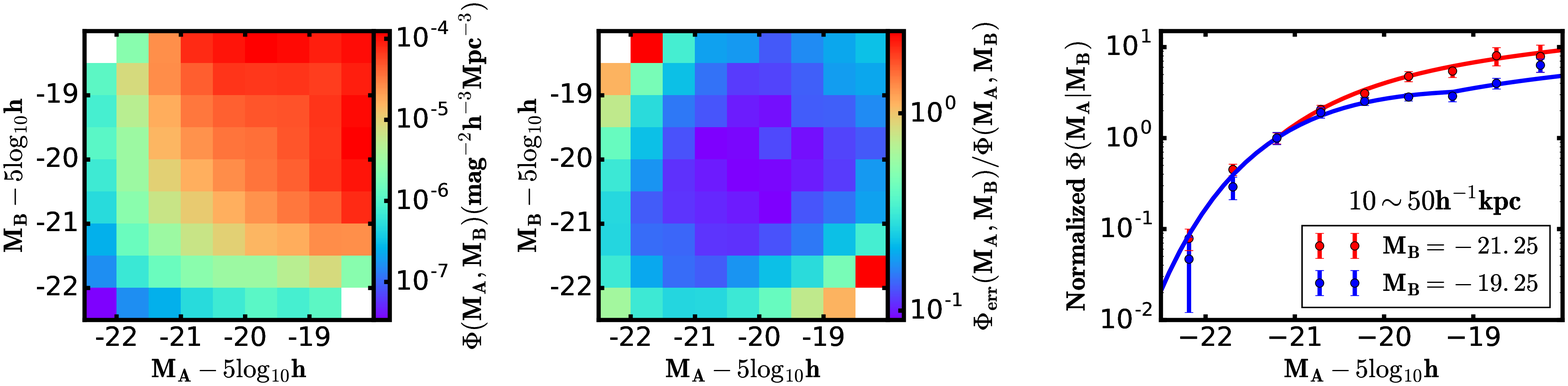}\\
\includegraphics[width=180mm]{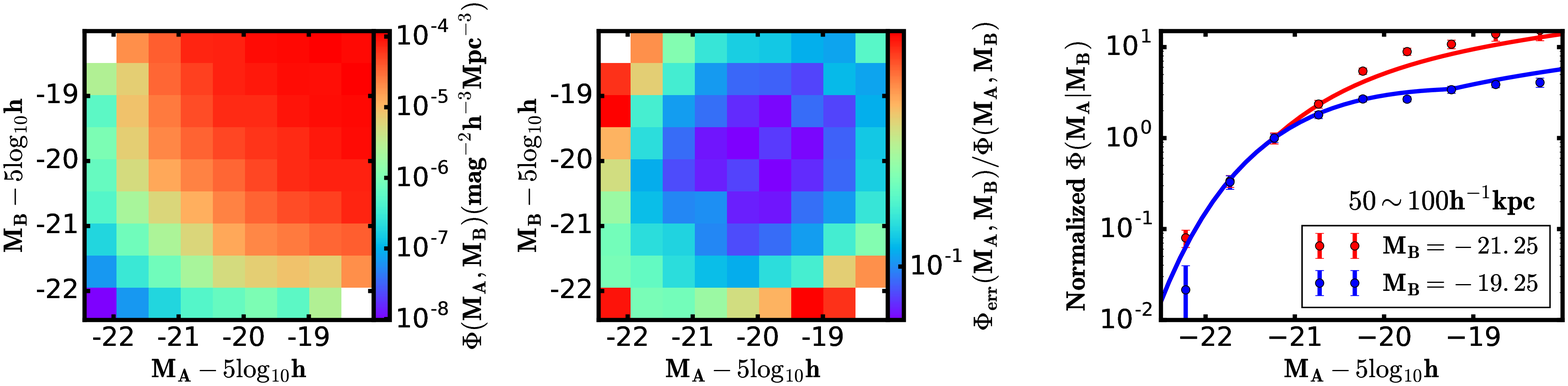}\\
\includegraphics[width=180mm]{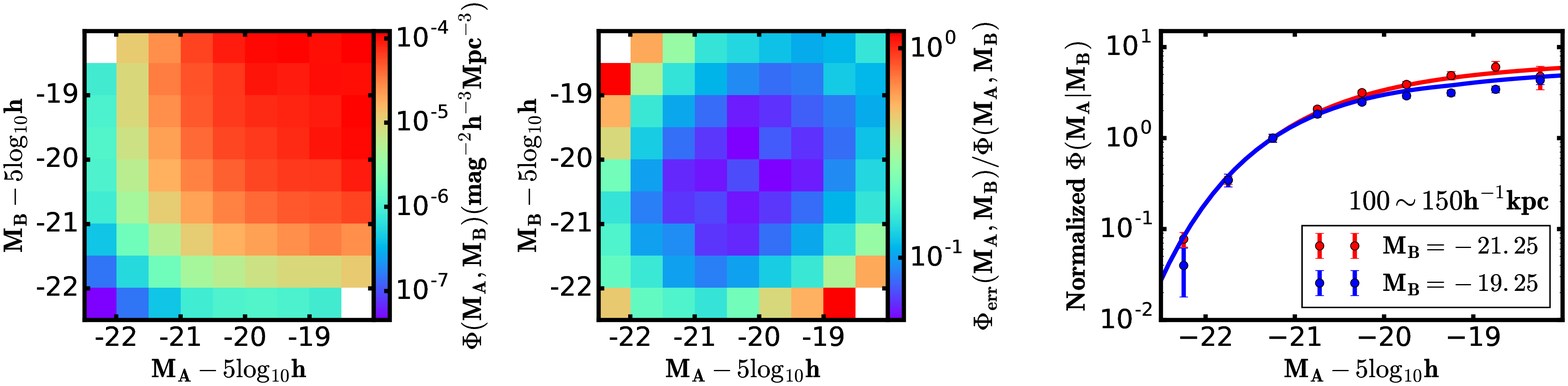}\\
\includegraphics[width=180mm]{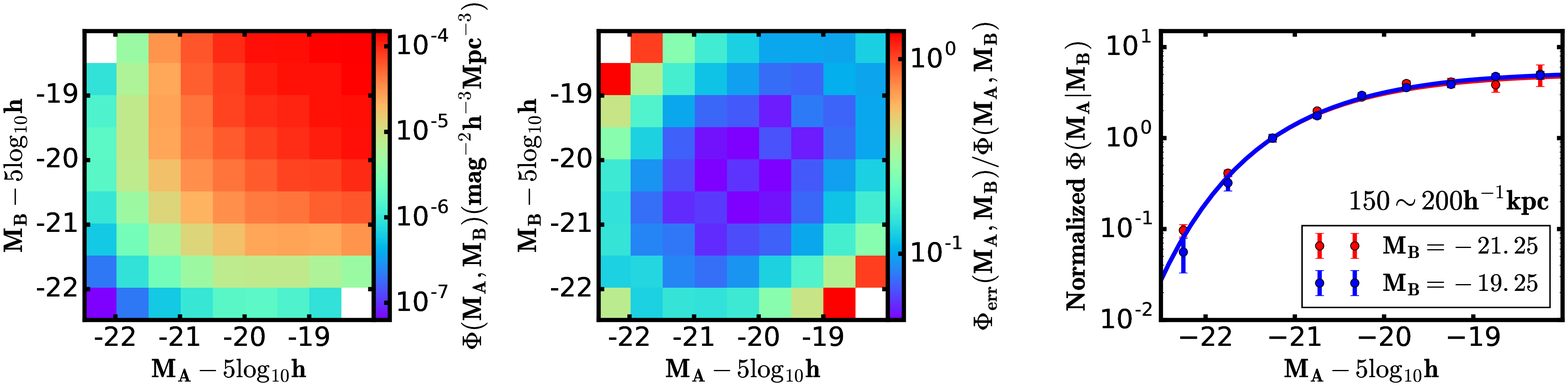}\\
\end{tabular}
\caption{BLFs for different $\pdis$ bins. \textit{Left panels: } BLF values. \textit{Middle panels:} relative errors of BLF. \textit{Right panels:}  the \textit{dots} with error bars indicate CLFs $\Phi(M_A|M_B=-21.25)$ and $\Phi(M_A|M_B=-19.25)$, whereas the \textit{solid lines} show the best fit of Equation \ref{eq_BLF}. }
\label{fig_clf}
\end{figure*}

We divide the galaxy pair sample into four subsamples according to the projected distances $\pdis$ between pair members and calculate the BLF of each subsample. The intervals of $\pdis$ are $10 \kpc \leq \pdis \leq 50 \kpc$, $50\kpc \leq \pdis \leq 100 \kpc$, $100\kpc \leq \pdis \leq 150 \kpc$ and $150\kpc \leq \pdis \leq 200 \kpc$ respectively. The BLFs are only calculated in the magnitude range of $-22.5 \leq M_{\textrm{r}} \leq -18.0$, where the number of galaxy pairs is large enough to make high confidence statistics. The resulting BLFs are shown in Figure \ref{fig_clf}, where four rows represent four subsamples. In each row, the left and middle panels show the BLF value and relative error respectively. To interpret the BLF more understandably, we show two CLFs $\Phi(M_A|M_B=-21.25)$ and $\Phi(M_A|M_B=-19.25)$ extracted from both the BLF map and its error map. The right panels show that when the projected distance increases to $\pdis \geq 150 \kpc$, $\Phi(M_A|M_B=-21.25)$ becomes indistinguishable from $\Phi(M_A|M_B=-19.25)$, indicating independence of two pair members. On the other hand, at $\pdis\leq 150 \kpc$, the CLFs start to show a discrepancy, and this discrepancy becomes even larger at smaller $\pdis$. 

The above BLF results explicitly show that the galaxies in pairs start to show interactions at a distance as far as $\pdis \sim 150 \kpc$. Except for a slight SFR enhancement out to $\pdis \sim 150 \kpc$ reported by \citet{patton2013}, most of the early studies conclude that the galaxies in pairs start to show significant interactions only at $\pdis \leq 50\kpc$ \citep{lambas2003,ellison2008,ellison2011,scudder2012,Davies2015}.The much larger critical $\pdis$ values revealed in this study is attributed to our statistical approach. Early studies typically compare the physical properties of galaxies in pairs with a control sample(those not in pairs). By control samples, the enhanced star formation in galaxy pairs could be revealed, but the galaxy merging effect, which changes the relative number density of pair members, could not be seen (see further discussion in Section \ref{sec_merge_time_scale}).

In general, at $\pdis \leq 150\kpc$, the CLFs of pair members with brighter neighbors [e.g., $\Phi(M_B|M_A=-21.25)$ ] are biased to distributions with a higher fraction of faint galaxies($M_B > -20.5$). It means, there are a lower fraction of bright-bright pairs than that expected from a random combination. This bias seems to be the strongest at $50 \kpc \leq \pdis \leq 100 \kpc$ rather than the smallest $\pdis$ bin ($10 \kpc \leq \pdis \leq 50 \kpc$). Such a complex behavior implies that different mechanisms dominate the galaxy-galaxy interaction within galaxy pairs at different phases (characterized by different $\pdis$ values), which will be discussed in Section \ref{sec_model}.

\subsection{Parametrization of BLF}\label{subsec_parameter}

The non-parametrized BLF maps shown in Figure \ref{fig_clf} are neither user-friendly nor easy to understand. Moreover, the CLFs extracted from BLF maps do not contain all the information of the BLFs. To overcome these shortcomings, we further attempt to approximate and parametrize the BLF maps using an analytical formula to describe better and quantify the BLF maps.

It is well known that the LF of galaxies can be well parametrized by a Schechter function \citep{schechter1976}, 
\begin{equation}\label{eq_schechter}
\begin{split}
\Phi(M)= & 0.4\ln(10)\phi^*10^{-0.4(M-M^*)(\alpha+1)} \\
& \times \exp[-10^{-0.4(M-M^*)}],
\end{split}
\end{equation}
where $\phi^*$ is the overall amplitude, $M^*$ and $\alpha$ are the characteristic magnitude and faint-end slope respectively.

Following this convention, we parametrize the BLF of galaxy pairs with,
\begin{equation}\label{eq_BLF}
\begin{split}
\Phi_{\textrm{pair}}(M_A,M_B) =& [0.4\ln(10)]^2\ \phi_{\textrm{pair}}^* \\
& \times \Phi_{\textmd{Sch}}(M_A) \Phi_{\textmd{Sch}}(M_B)X(M_A,M_B),
\end{split}
\end{equation}
where $\Phi_{\textmd{Sch}}(M)$ is a Schechter-like function representing the LF shape of single galaxies, and the term $X(M_A,M_B)$ describes the correlation between pair members. More specifically, $\Phi_{\textmd{Sch}}(M)$ takes the form of Equation \ref{eq_schechter} but without the density parameter $\phi^*$. The density parameter of galaxy pairs is represented by $\phi_{\textrm{pair}}^*$ in Equation \ref{eq_BLF}. 

We parameterize the correlation term $X(M_A,M_B)$ with an assumed form,
\begin{equation}
X(M_A,M_B)=e^{\beta|M_A-M_B|},
\end{equation}
where $\beta$ is the only parameter quantifying the strength of correlation.

For the case of $\beta=0$, in which $\Phi_{\textrm{pair}}(M_A,M_B) \propto\Phi_{\textmd{Sch}}(M_A) \Phi_{\textmd{Sch}}(M_B)$, there is no correlation between pair members and the BLF degenerates into the product of the LF of single galaxies.  In other words, the galaxy pairs are built through a random combination of single galaxies. When $\beta >0$, the galaxy pairs with large magnitude gap ($|M_A-M_B|$) account for a larger fraction than a random combination. On the contrary, $\beta < 0$ means that the fraction of galaxy pairs with large magnitude gaps is lower. With such a parametrization, we would expect that $\beta \approx 0 $ at $\pdis > 150 \kpc$. For the other three $\pdis$ intervals, we expect $\beta>0$, since we have shown in Figure \ref{fig_clf} that the brighter galaxies($M_A=-21.25$) have more faint companions than random combinations, i.e., the galaxy pairs with a larger magnitude gap are more preferred.

We fit BLFs of all sub-samples with Equation \ref{eq_BLF} using the minimum $\chi^2$ fitting technique. We estimate the uncertainty of each fitting parameter using  the $16$th and $84$th percentiles (half of the interval) of the marginal distribution of the likelihood, which is defined as
\begin{equation}
\mathscr{L}=\exp(-\frac{\chi ^2}{2}) \,.
\end{equation}

First, we fit the BLF of the galaxy pairs within the bin $150 \kpc \leq \pdis \leq 200 \kpc$, where $\alpha, M^*, \phi^*_{\text{pair}}$ and $\beta$ are all set as free parameters. As expected, the best fit has $\beta=0.00\pm0.02$, confirming our conjecture that there is no correlation between the pair members within the bin $150 \kpc \leq \pdis \leq 200\kpc$. The best fit of the other three parameters are $\phi^*_{\text{pair}}=2.12 \pm 0.04 \times 10^{-4} \text{h}^{-3}\ \text{Mpc}^{-3}$, $M^*=-20.71 \pm 0.02$,$\alpha=-1.01\pm 0.02$ respectively. Next, for the galaxy pairs in the other three $\pdis$ intervals, we fix $\alpha=-1.01$ and fit the BLF with $M^*$,$\phi^*_{\text{pair}}$ and $\beta$ as free parameters. Since the lower magnitude limit of pair members is only down to $-18$ mag, if $\alpha$ is also set as a free parameter, it would show strong degeneracy with the characteristic magnitude $M^*$. Therefore, if we fix $\alpha$, the change in BLF shape with $\pdis$ could be reproduced by the variation of $M^*$ and $\beta$, which could also be directly compareted to the BLF within the bin $150 \kpc \leq \pdis \leq 200 \kpc$.

\begin{table*}
\centering
\caption{The best fitting parameters and their uncertainties in Equation \ref{eq_BLF} for the BLFs in four different $\pdis$ intervals.}
\label{tb_blf_fitting}
\begin{tabular}{cccccccc}
\hline
$\pdis$ & $N_{\textmd{pair}}$ & $\phi^*_{\text{pair}}$ & $M^*$ & $\alpha$ & $\beta$ & $\chi^2_{\text{dof}}$ \\
$(\textmd{kpc})$ &   & $(10^{-4}\ \text{h}^{-3}\ \text{Mpc}^{-3})$ &  &  &  &  &  \\
\hline
$10 \sim 50 $ & 6826  & $1.26 \pm 0.04$ & $-20.58 \pm 0.02$ & $-$ & $0.21 \pm 0.03$ & 1.73\\
$50 \sim 100$ & 11314 & $1.78 \pm 0.05$ & $-20.48 \pm 0.01$ & $-$ & $0.24 \pm 0.02$ & 1.01\\
$100\sim 150$ & 13657 & $1.86 \pm 0.04$ & $-20.68 \pm 0.01$ & $-$ & $0.05 \pm 0.02$ & 0.99\\
$150\sim 200$ & 14713 & $2.12 \pm 0.04$ & $-20.71 \pm 0.02$ & $-1.01\pm 0.02$ & $0.00 \pm 0.02$ & 0.79\\
\hline
\end{tabular}
\end{table*}

The best fitting parameters of all four BLF maps are listed in Table \ref{tb_blf_fitting}. All four bins have $\chi^2_{\text{dof}}\sim 1$, indicating that the parametrized BLFs (Equation \ref{eq_BLF}) can well represent the non-parametrized BLF maps shown in Figure \ref{fig_clf}. As an illustration of the goodness of fit, we also plot two CLFs, $\Phi(M_B|M_A=-21.25)$ and $\Phi(M_B|M_A=-19.25)$, deduced from the best fitting parameters, as solid lines in the right panels of Figure \ref{fig_clf} for comparison.

\begin{figure*}
\centering
\includegraphics[width=180mm]{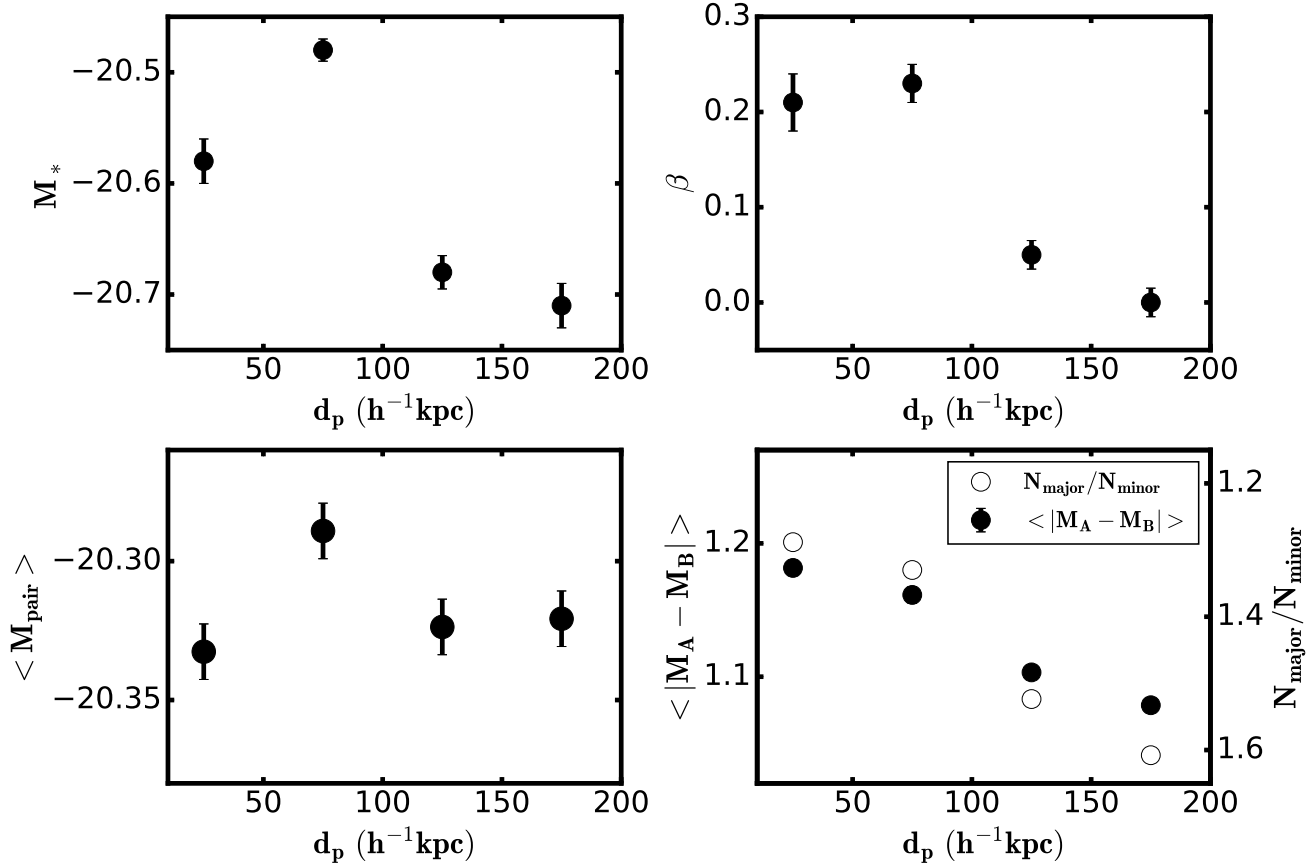}
\caption{Parameterization of BLF variation. \textit{Top panels:} the best fitting values in Equation \ref{eq_BLF} when $\alpha$ is fixed, including the character magnitude $M^*$ (\textit{top left}) and correlation strength $\beta$ (\textit{top right}). \textit{Bottom panels:} the statistical quantities of our galaxy pair sample, including the mean total magnitude of the galaxy pairs $\langle M_{\text{pair}} \rangle$ (\textit{bottom left}), mean magnitude gap of the galaxy pairs $\langle |M_{\text{A}}-M_{\text{B}}| \rangle$ and number density ratio of the major merger pairs to minor merger pairs $N_{\text{major}}/N_{\text{minor}}$(\textit{bottom right}). In the bottom right panel, the values of  $\langle |M_{\text{A}}-M_{\text{B}}| \rangle$ and  $N_{\text{major}}/N_{\text{minor}}$ are labelled on the left and right $y$-axis respectively. }
\label{fig_blf_pdis_exp}
\end{figure*}

We show the best fitting parameters $\beta$ and $M^*$ as functions of $\pdis$ in the top two panels of Figure \ref{fig_blf_pdis_exp}. With the decrease of $\pdis$, $\beta$ shows an increasing trend until $\pdis$ reaches $50 \kpc$, then stays unchanged. The variation of $M^*$ is similar, which gradually increases (becomes fainter) with decreasing $\pdis$ and then drops (becomes brighter) within the innermost bin ($10 \kpc \leq \pdis \leq 50 \kpc$). The increasing trend of $\beta$ implies that the galaxy-galaxy interaction gets stronger when they get closer. Because of the correlation term $\beta$, we remind that the physical implication of $M^*$ is not as straightforward as that in the LF of single galaxies. Therefore, the variation of the best fitting parameters listed in Table \ref{tb_blf_fitting} cannot be over-interpreted. We will discuss the physical implications of these BLF results in more details in Section \ref{sec_model}. 

\subsection{Univariate Luminosity Function and Number Density of Galaxy Pairs}

From the BLF shown in Figure \ref{fig_clf}, we can obtain the ULF of galaxy pairs through integrating the magnitudes of one member, 
\begin{equation}
\Phi(M_A) =\int_{-22.5}^{-18.0} \Phi_{\textrm{pair}}(M_A,M_B)dM_B  \,.
\end{equation}
Here, the integration can only be performed in the magnitude range where the BLF has been calculated. Namely, the galaxy pairs here have been strictly defined as those both of whose members are in the magnitude range of $-22.5 \leq M_r \leq -18.0$. 

\begin{figure}
    \centering
    \includegraphics[width=90mm]{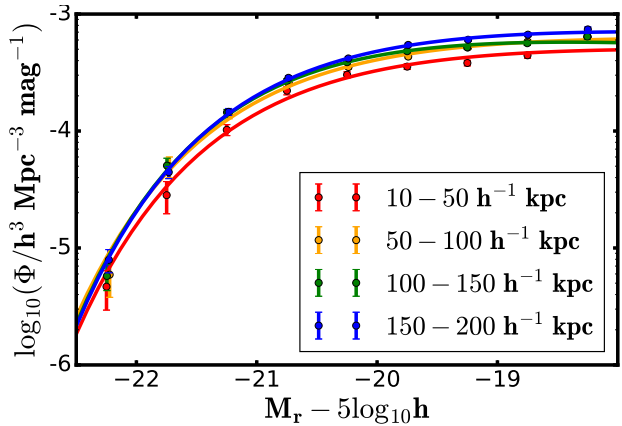}
    \caption{The ULFs of four subsamples. \textit{Dots:} the ULFs derived from BLFs. \textit{Solid Lines:} the best fits by Schechter Function. }
    \label{fig:SLF}
\end{figure}

In Figure \ref{fig:SLF}, we show the resulting ULFs of the galaxy pairs within four different $\pdis$ bins. We also fit these ULFs with a Schechter function, and the resultant fittings are shown as the corresponding solid curves in Figure \ref{fig:SLF}. The fitting parameters of the Schechter function are also listed in Table \ref{tab:ulf}. Since our galaxy pair sample is a subsample of the SDSS MGS, the ULFs can be directly compared with the global LF of SDSS galaxies \citep{blanton2003a}.  Because the galaxy pair is one kind of environments for a galaxy,  the ULF of galaxy pairs is expected to be different from the global LF of galaxies. This is indeed the case. For example, for the galaxy pairs within the widest $\pdis$ bin ($150 \kpc \leq \pdis \leq 200 \kpc$), where even none galaxy-galaxy interaction effects have been detected in the BLF measurement, their ULF still shows a brighter characteristic magnitude $-20.52$ and a shallower faint-end slope $-0.92$ than the general SDSS galaxies($M_*=-20.44,\alpha=-1.04$).  Such behavior is consistent with the earlier findings for the general trend of the environmental dependence of the LF of galaxies \citep[e.g.][]{McNaught-Roberts2014,alpaslan2015}.

The shapes of the ULFs within four different $\pdis$ bins are quite similar with values of $M^*$ and $\alpha$ overlapping each other inside 1-$\sigma$ uncertainties. The similarities of the ULFs contrast with the BLF behaviors,  which reveals different physical mechanisms behind two types of LFs.  Generally, BLFs describe the galaxy interactions between pair members, while the ULFs are further correlated with the global environment of galaxy pairs.   The weak dependence of the ULFs on $\pdis$ implies that the global environments of different types of galaxy pairs are quite similar.

\begin{table}[]
    \centering
    \caption{The best fitting parameters and their uncertainties of the ULFs in four different $\pdis$ intervals.}
    \begin{tabular}{cccc}
    \hline
    $\pdis$ & $\phi^*$ & $M^*$ & $\alpha$ \\
    $(\textmd{kpc})$ & $(10^{-4}\ \text{h}^{-3}\ \text{Mpc}^{-3})$ &  &   \\
    \hline
    $10 \sim 50 $ & $6.59 \pm 0.06$ & $-20.58 \pm 0.08$ & $-0.95\pm 0.06$ \\
    $50 \sim 100$ & $8.01 \pm 0.06$ & $-20.61 \pm 0.07$ & $-0.96\pm 0.05$ \\
    $100\sim 150$ & $9.57 \pm 0.05$ & $-20.51 \pm 0.05$ & $-0.85\pm 0.04$ \\
    $150\sim 200$ & $10.73 \pm 0.06$ & $-20.52 \pm 0.05$ & $-0.92\pm 0.04$ \\
    \hline
    \end{tabular}
    \label{tab:ulf}
\end{table}

The amplitude of the BLFs/ULFs tells us the number density of galaxy pairs. As shown by $\phi^*_{\text{pair}}$ and $\phi^*$ in Table \ref{tb_blf_fitting} and \ref{tab:ulf}, the global number densities of paired galaxies in the SDSS volume are quite similar within four different $\pdis$ intervals. These characteristics of the global number densities have already been seen in the flat $\pdis$ distribution of our galaxy pair sample after the correction of incompleteness(red dashed line in the bottom left panel of Figure \ref{fig_samp}). On the other hand, since the local volumes occupied by galaxy pair themselves are smaller for smaller $\pdis$ values, the resulting local number density of the galaxies in close pairs is higher than that in wide pairs. The different behaviors of the global and local number densities of paired galaxies exhibit the galaxy clustering effect, where galaxies are more clustered in smaller $\pdis$ scales.\footnote{Because of the limited number of galaxies and complicated selection effects, the studies of galaxy clustering effects at very small scales(e.g., $\pdis < 100\kpc$) are very limited \citep[e.g.][]{guo2012}. }

By integrating the BLFs, we can further estimate the number densities of galaxy pairs with any specific magnitude configuration. For example, we have estimated from the BLF map that the number density of galaxies in the range of $-19 \leq M_r \leq -21.5$ and locating in the the galaxy pairs with $\Delta M_r \leq 1$ and $10 \kpc \leq \pdis \leq 50 \kpc$ equals to $n= 4.59 \times 10^{-3}\ \text{h}^{-3}\ \text{Mpc}^{-3}$. Accounting the number density of the general SDSS galaxies in the same magnitude range($\sim 0.015 \text{h}^{-3}\ \text{Mpc}^{-3}$ from \citealt{blanton2003b}), we conclude that $3.06\%$ of the SDSS galaxies locate in galaxy pairs with $\Delta M_r <1$ and $10 \kpc \leq \pdis \leq 50 \kpc$ . This result is consistent with the early estimation of the pair fraction from the ULF of galaxy pairs ($\sim 1.6\%$ \citealt{domingue2009}), where the galaxy pairs are limited to those with $\Delta M_{K_S}<1$ and $5 \kpc \leq \pdis \leq 20 \kpc$.

\section{Physical interpretation of the BLF Variation}\label{sec_model}

In this section, we first quantify the BLF dependence on $\pdis$ by defining two new statistical quantities (Section \ref{sec_BLFquantity}) and then discuss their variation with physically motivated toy models(Section \ref{sec_merge_time_scale} and \ref{sec_starformation}).

\subsection{$\langle M_{\text{pair}} \rangle$ and $\langle |M_{\text{A}}-M_{\text{B}}| \rangle$ of Galaxy Pairs}\label{sec_BLFquantity}

For individual galaxy pairs, the total magnitude $M_{\text{pair}}$ and magnitude gap $|M_{\text{A}}-M_{\text{B}}|$ are sufficient to fully describe the magnitude configuration of its members. For statistical samples, the mean values of these two quantities, $\langle |M_{\text{A}}-M_{\text{B}}| \rangle$ and $\langle M_{\text{pair}} \rangle$ present a first order description of the BLF shape. Specifically, $\langle M_{\text{pair}} \rangle$ describes the average total magnitude of pairs, whereas $\langle |M_{\text{A}}-M_{\text{B}}| \rangle$ mainly reflects the magnitude configuration. In addition, $|M_{\text{A}}-M_{\text{B}}|$ is also used to define the merging type of each galaxy pair, if the mass-to-light ratio of pair members are assumed to be the same. Following convention, we define the major and minor merger pairs according to the criteria $1/3 \leq L_{\text{A}}/L_{\text{B}} \leq 3$. Typically, larger $\langle |M_{\text{A}}-M_{\text{B}}| \rangle$ indicate a lower fraction of major merger pairs in the galaxy pair sample.

We calculate the average total magnitude $\langle M_{\text{pair}} \rangle$ and the average magnitude gap $\langle |M_{\text{A}}-M_{\text{B}}| \rangle$ of each subsample from the BLF maps. We show $\langle M_{\text{pair}} \rangle$ and $\langle |M_{\text{A}}-M_{\text{B}}| \rangle$ as functions of $\pdis$ in lower two panels of Figure \ref{fig_blf_pdis_exp}, where the number density ratio of the major merger pairs to minor merger pairs $N_{\text{major}}/N_{\text{minor}}$ is also incorporated in the bottom right panel.

As $\pdis$ decreases, $\langle |M_{\text{A}}-M_{\text{B}}| \rangle$ increases ($N_{\text{major}}/N_{\text{minor}}$ decreases) monotonously. Such a behavior is very similar to the trend of $\beta$  shown in the top right panel.  By contrast, the variation of $\langle M_{\text{pair}} \rangle$ is similar to the trend of $M^*$, which increases gradually and then drops within the innermost $\pdis$ bin.

The variation of the BLFs and the BLF parameters as functions of $\pdis$ shown in Figure \ref{fig_clf} and \ref{fig_blf_pdis_exp} implies that there are three \textit{phases} during the galaxy pair evolution. In the beginning, i.e., the first phase, when the distance of two pair members is very large($\pdis \geq 150 \kpc$), there is little interaction between them, and the BLF of galaxy pairs degenerates into the LF of single galaxies (bottom panels of Figure \ref{fig_clf}). In the second phase, when $50 \kpc \leq \pdis \leq 150 \kpc$, both the fraction of the major merger pairs and the average total luminosity of galaxy pairs systematically decrease with decreasing $\pdis$. These results imply that the interaction drives the number of luminous major merger pairs successively to drop. This behavior resembles the merging process induced by dynamic friction where the merging time-scale of the massive major merger pairs is shorter than others (see further discussion in Section \ref{sec_merge_time_scale}). For the third phase ($10 \kpc \leq \pdis \leq 50 \kpc$), the effect of dynamic friction still remains ($\langle |M_{\text{A}}-M_{\text{B}}| \rangle$ keeps increasing), whereas another effect in turn brightens the average total luminosity of galaxy pairs. In close pairs (e.g. $\pdis < 50\kpc$), it has long been known that the close interaction enhances the star formation so that also brightens the luminosity of the paired galaxies(\citealt{ellison2008,scudder2012,patton2013,Davies2015}, see further discussion in Section \ref{sec_starformation}).

\subsection{Merging Time-scale}\label{sec_merge_time_scale}

Once the galaxy pairs are in bound orbits, the dynamical friction will drive the pair members to approach each other and finally merge. The detailed merging process of a specific galaxy pair includes a very complex chain of events, and the merging time-scale is related to many details, e.g., the galaxy orbit, mass, angular momentum and internal structure of the paired galaxies etc. \citep{colpi1999,Boylan-Kolchin2000,lotz2010b,lotz2010a} However, numerical simulations have shown that the merging time-scale of galaxy pairs mainly depends on the masses of pair members, and the typical merging time-scale (from a bound orbit to the final merge) is about $\sim 1-2$ Gyr (e.g., \citealt{Jiang2014}).

Many studies have shown that galaxy pairs with comparable stellar masses (i.e., major merger pairs), merge more quickly than the minor merger ones(e.g. \citealt{Boylan-Kolchin2000}). This effect is typically parameterized by
\begin{equation*}
T_{\text{m}} \sim (\frac{M_{h,1}}{M_{h,2}})^a
\end{equation*}
where $M_{h,1}$ and $M_{h,2}$ are the halo masses of the primary and secondary galaxies respectively. Although the consensus is that $a > 0$, the index $a$ spans a wide range ( from $\sim 0.4$ to $\sim1.3$) in different studies \citep{lacey1994,Colpi1997,Boylan-Kolchin2000}. Besides the mass ratio $M_1/M_2$, the merging time-scale is also correlated with the mass of primary galaxy  \citep{Jiang2014}. Namely, a more massive galaxy pair would also have a shorter $T_{\text{m}}$($T_{\text{m}} \sim M_{h,1}^{-1/3}$ in \citealt{Jiang2014}).

Taking all these factors into account, and assuming a constant dark matter halo mass-to-light ratio of galaxies, we parameterize $T_{\text{m}}$ as a function of the luminosities of pair members,
\begin{equation}\label{eq_tdy}
T_{\textrm{m}}(L_{\text{A}},L_{\text{B}}) \sim T_{\textrm{m}}(L_{1},L_{2}) \sim (\frac{L_1}{L_2})^a (L_1+L_2)^b \,
\end{equation}
where $L_1$ and $L_2$  represent the luminosities of the primary and secondary members, $a$ and $b$ are free parameters. 

In a hierarchical clustering universe, galaxy pairs bound and merge continuously. We assume that galaxy pairs are formed through a random combination of field/single galaxies when their two members approach each other at a critical distance (e.g., $\pdis \sim 175 \kpc$). Thus, the birth rate of galaxy pairs is proportional to the product of the number densities of two pair members, 
\begin{equation}\label{eq_birthrate}
\dot{\Phi}({L_{\text{A}},L_{\text{B}}}) \propto \Phi_0(L_{\text{A}})\Phi_0(L_{\text{B}}) \,
\end{equation}
where $\Phi_0$ is the LF of single/field galaxies. As we have shown in Section \ref{subsec_parameter}, the galaxy pairs within the bin $150\kpc \leq \pdis \leq 200\kpc$ show no internal correlation and can be viewed as random combinations of single galaxies. The excellent fitting of Equation \ref{eq_BLF} for the BLF of galaxy pairs within this $\pdis$ interval tells us that the LF of single galaxies $\Phi_0$ to be forming galaxy pairs could be parameterized by a Schechter function with $M^*=-20.73$ and $\alpha=-1.01$.

Then, the observed number density of galaxy pairs with a given configuration is further proportional to their lifespan, i.e., the merging time-scale $T_{\textrm{m}}$,
\begin{equation}\label{eq_N}
N(L_{\text{A}},L_{\text{B}}) \propto \Phi_0(L_{\text{A}})\Phi_0(L_{\text{B}})T_{\textrm{m}}(L_{\text{A}},L_{\text{B}}) \,.
\end{equation}

These galaxy pairs would span a wide distribution of projected distance $\pdis$, ranging from $d_{\text{min}}$ to $d_{\text{max}}$, where $d_{\text{min}}$ and $d_{\text{max}}$ are the minimum and maximum projected distances between pair members. We denote the number density of galaxy pairs at a certain $\pdis$ as $N(L_{\text{A}},L_{\text{B}},\pdis)$, and assume that the $N(L_{\text{A}},L_{\text{B}},\pdis)$ distribution follows a linear relation with $\pdis$. Namely, once we have constraints on the number density of galaxy pairs at both $d_{\text{min}}$ and $d_{\text{max}}$, the $N(L_{\text{A}},L_{\text{B}},\pdis)$ distribution could be easily figured out.

\begin{figure}
\centering
\includegraphics[width=0.5\textwidth]{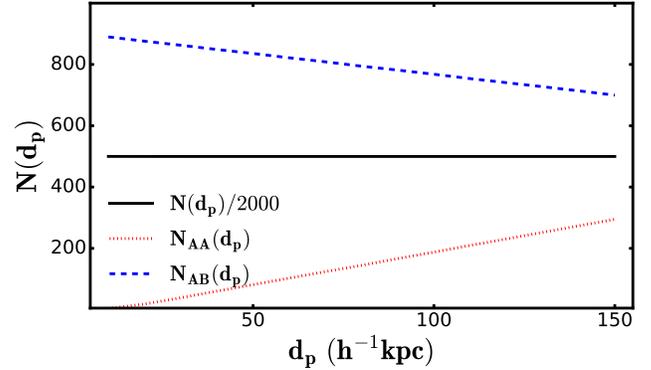}
\caption{The $\pdis$ distribution of mock galaxy pairs generated from Monte-Carlo simulation(see the text in Section \ref{sec_merge_time_scale} for details).The \textit{black solid line} shows the $\pdis$ distribution of all 15,000,000 mock pairs, where the y-axis values have been divided by 200 for clarity. The \textit{red dotted} and \textit{blue dashed} line show the $\pdis$ distribution of galaxy pair AA $(-20.75\pm0.25,-20.75\pm0.25)$ and AB $(-20.75\pm0.25,-18.75\pm0.25)$ respectively.}
\label{fig_ladder_plot}
\end{figure}

For the galaxy pairs at $d_{\text{max}}$, we know that their relative number density is independent of the merging time-scale and is only determined by their birth rate, i.e.,
\begin{equation}\label{eq_ndmax}
N(L_{\text{A}},L_{\text{B}},d_{\text{max}}) \propto \dot{\Phi}_{L_{\text{A}},L_{\text{B}}} \propto \Phi_0(L_{\text{A}})*\Phi_0(L_{\text{B}}) \,.
\end{equation}
Moreover, with the key assumption of a linear $N(L_{\text{A}},L_{\text{B}},\pdis)$ distribution,  the number density of galaxy pairs at $d_{\text{min}}$, $N(L_A,L_B,d_{\text{min}})$ could be estimated through the relation
\begin{equation}\label{eq_nn}
N(L_{\text{A}},L_{\text{B}})=0.5\frac{N(L_{\text{A}},L_{\text{B}},d_{\text{max}})+N(L_{\text{A}},L_{\text{B}},d_{\text{min}})}{(d_{\text{max}}-d_{\text{min}})} \,.
\end{equation}
So far, the last piece of the $N(L_{\text{A}},L_{\text{B}},\pdis)$ distribution is the normalization constant of Equation \ref{eq_N} and \ref{eq_ndmax}, which also determines the global $\pdis$ distribution of galaxy pairs. Motivated by our galaxy pair sample, a reasonable assumption on the global $\pdis$ distribution is a flat distribution (see the red dotted line in the bottom left panel of Figure \ref{fig_samp}).

In practice, for a given parameterization of merging time-scale(Equation \ref{eq_tdy}), we use a Monte Carlo simulation to generate 15,000,000 mock galaxy pairs using Equation \ref{eq_tdy} and \ref{eq_N}, where $\Phi_0$ takes the Schechter function with $M^*=-20.73$ and $\alpha=-1.01$. As for the observed sample, the mock galaxies are set in the magnitude range of $-18.0$ to $-22.5$ mag and $\pdis$ is set in the range of $d_{\text{min}}=10\kpc$ to $d_{\text{max}}=150\kpc$. We divide the mock pairs into $15$ $\pdis$ bins with a bin width of $10\kpc$. Since the mock galaxy pair sample has been assumed to have a flat $\pdis$ distribution, the number of mock pairs within each $\pdis$ bin is therefore $\sim 1,000,000$. For galaxy pairs with specific configuration $(L_A,L_B)$, their numbers within the $d_{\text{max}}$ bin ($145-155\kpc$) follow the distribution of a random combination of single galaxies as shown by Equation \ref{eq_ndmax}. Then, the number of the galaxy pairs within the $d_\text{min}$ bin is calculated using  Equation \ref{eq_nn}.  Finally, the number of mock pairs within any given $\pdis$ bin could be easily interpolated from $N(L_A,L_B,d_\text{max})$ and $N(L_A,L_B,d_\text{min})$.

To give an example, we show the mock galaxy pairs with two specific magnitude configuration $(-20.75\pm0.25,-20.75\pm0.25)$ and $(-20.75\pm0.25,-18.75\pm0.25)$. These represent the major and minor merger pairs respectively and are denoted as AA and AB below. Here, we take the modeling for the merging time-scale(Equation \ref{eq_tdy}) of $a=0.40$ and $b=-0.33$ as an illustration.  First, among $15,000,000$ mock pairs, there are $2,264$ AA and $12,126$ AB respectively(Equation \ref{eq_N}). Then, with Equation \ref{eq_ndmax}, we know that there are $302$ AA and $713$ AB within the $d_{\text{max}}$ bin. Using Equation \ref{eq_nn}, we obtain that there are $4$ AA and $904$ AB within the $d_{\text{min}}$ bin. Finally, the numbers of AA and AB within all $\pdis$ bins are obtained from linear interpretation, which is illustrated in Figure \ref{fig_ladder_plot}.  As can be seen, because of their different merging time-scales, AA (major merger pairs) show a significant different $\pdis$ distribution from AB (minor merger pairs). In our modeling, because of their longer merging time-scale, there is an increasing fraction of minor merger pair in inner $\pdis$ regions. This phenomenon is consistent with a scenario, that the minor merger pairs experience more orbits before coalescing and are more frequently observed in the inner $\pdis$ regions \citep{vdb1999}. 

\begin{figure*}
\centering
\includegraphics[width=180mm]{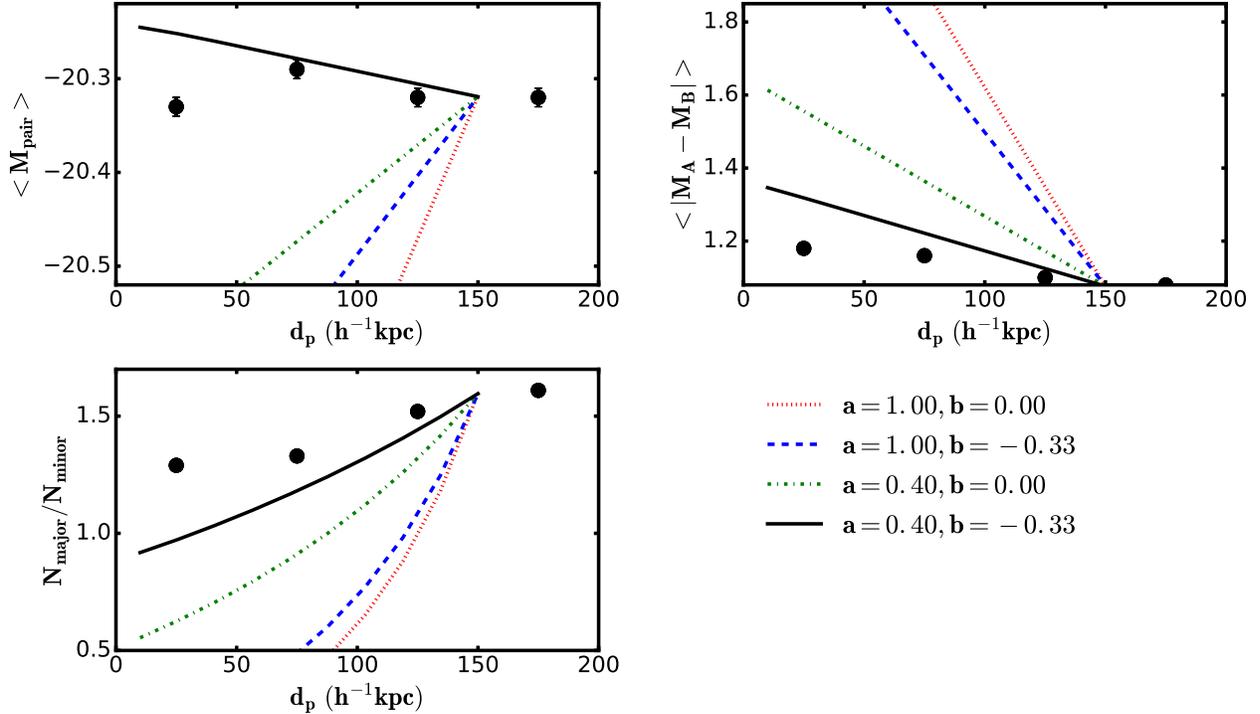}
\caption{Modelling of the BLF variation with $\pdis$. \textit{Black dots:} the observational values of the BLF parameters, including the mean total magnitude (\textit{top left}), mean magnitude gap (\textit{top right}) and number density ratio between major merger pairs and minor merger pairs (\textit{bottom left}). \textit{Lines:} BLF variation predicted by the merging time-scale model(Section \ref{sec_merge_time_scale}), where different lines indicate different merging time-scales parameterized by Equation \ref{eq_tdy}. }
\label{fig_model_dynfric}
\end{figure*}

The toy modeling we presented above may not be enough to characterize all the variations of BLFs. Therefore, we do not attempt to model the variation of the exact shape of BLF but to focus on the general trends of the BLF parameters as functions of $\pdis$. Specifically, we model the trends of $\langle |M_{\text{A}}-M_{\text{B}}| \rangle$ , $\langle M_{\text{pair}} \rangle$ and $N_{\text{major}}/N_{\text{minor}}$, then compare them with the observational results already shown in Section \ref{sec_BLFquantity}.

We first take two sets of parameterizations of $T_{\textrm{m}}$  from \citet{lacey1994,Colpi1997}, i.e., ($a=1.0,b=0$) and ($a=0.4,b=0$). The results are plotted as red dotted and green dot-dashed lines in each panel of Figure \ref{fig_model_dynfric}. In these two models, luminosity/mass ratio is the only component of $T_{\textrm{m}}$. A positive $a$ means that the merging time-scale of minor merger pairs is larger than the major merger ones, which naturally  results in a larger average $\langle |M_{\text{A}}-M_{\text{B}}| \rangle$ and a higher fraction of minor merger pairs. Moreover, both two models predict the trend of increasing $\langle |M_{\text{A}}-M_{\text{B}}| \rangle$ (decreasing $N_{\text{major}}/N_{\text{minor}}$) with the decreasing $\pdis$. This trend is steeper for larger $a$ values.  Moreover, in the $a>0$ models, the pairs with very bright primary galaxies are also preferred (because they are mainly minor merge pairs), which in turn results in a brightening of $\langle M_{\text{pair}} \rangle$. This decreasing trend of $\langle M_{\text{pair}} \rangle$ with decreasing $\pdis$ is in conflict with observations. The results shown above indicate that although smaller $a$ is preferred by the observations, this simple parameterization is not enough to explain all the observed BLF trends. 

Next, we take both the luminosity/mass ratio and total luminosity/mass into account. Following \citet{Jiang2014},  we take  $a=1.0,b=-0.33$ and the results are shown as the blue dashed line in Figure \ref{fig_model_dynfric}. When $b<0$, the merging time-scale of  brighter galaxy pairs is smaller, which results in a fainter $\langle M_{\text{pair}} \rangle$ than the corresponding $b=0$ model. However, in this model, the effect of $b=-0.33$ seems not enough to balance the effect of $a=1.0$ on the  predicted $\langle M_{\text{pair}} \rangle$ trend.

So far, all these three models inferred from literature could not reproduce the general trends of $\langle M_{\text{pair}} \rangle$ and $\langle |M_{\text{A}}-M_{\text{B}}| \rangle$ simultaneously. A natural conclusion from comparison of these three models is that a smaller positive $a$ value and negative $b$ value would be more consistent with observations, which leads to our final test model with $a=0.4$ and $b=-0.33$. The model predictions are shown as the black solid lines in Figure \ref{fig_model_dynfric}.  This model reproduces all the global trends of $\langle |M_{\text{A}}-M_{\text{B}}| \rangle$ and $N_{\text{major}}/N_{\text{minor}}$. For $\langle M_{\text{pair}} \rangle$, this model also  predicts the general trend well, except the smallest $\pdis$ bin where a jump(brightening) of $\langle M_{\text{pair}} \rangle$ is clearly seen.

We emphasize that the fourth model with $a=0.4$ and $b=-0.33$ is only a test and not a fitting of the observed trends of $\langle M_{\text{pair}} \rangle$ and $\langle |M_{\text{A}}-M_{\text{B}}| \rangle$. As already implied from four test models, there is certain degeneracy of model predictions between the model parameters $a$ and $b$. The discrepancy between the observed and modeled trends of $\langle M_{\text{pair}} \rangle$ and $\langle |M_{\text{A}}-M_{\text{B}}| \rangle$ could be further alleviated by taking either a smaller positive $a$ value or a more negative $b$ value. Nevertheless, none of the parameters ($a$ or $b$) alone can  reproduce the observed $\langle M_{\text{pair}} \rangle$ and $\langle |M_{\text{A}}-M_{\text{B}}| \rangle$ trends simultaneously.

\subsection{Enhanced Star Formation in Close-Pair Phase}\label{sec_starformation}

As we have shown in Section \ref{sec_merge_time_scale}, the merging time-scale effect could well explain most of the BLF variation features, except that the average total luminosity of galaxy pairs within the bin $10\kpc \leq \pdis \leq 50 \kpc$ becomes significantly brighter. Such a feature is consistent with the scenario that there is enhanced star formation in close pairs \citep{ellison2008}, which has not been taken into account so far. In this section, we discuss how star formation enhancement induces galaxy brightening, and thus changes the BLF of galaxy pairs. For the sake of brevity, we name the galaxy pairs within the bin $10\kpc \leq \pdis \leq 50 \kpc$ as close pairs below.

Quantitatively, our BLF measurements show that the average total magnitude ($\langle M_{\text{pair}} \rangle$) of close pairs is $0.043$ mag brighter than the galaxy pairs within the bin $50\kpc \leq \pdis \leq 100\kpc$. However, besides the enhanced star formation effect, the magnitude configurations of close pairs may have been further altered by the merging time-scale effect compared to those within the bin $50\kpc \leq \pdis \leq 100\kpc$ (see the solid line in the top left panel of Figure \ref{fig_model_dynfric}).  Therefore,  the average brightening of the galaxies in the close-pair phase, $\Delta M_r \sim 0.04$, is quite conservative.

\begin{figure}
\includegraphics[width=90mm]{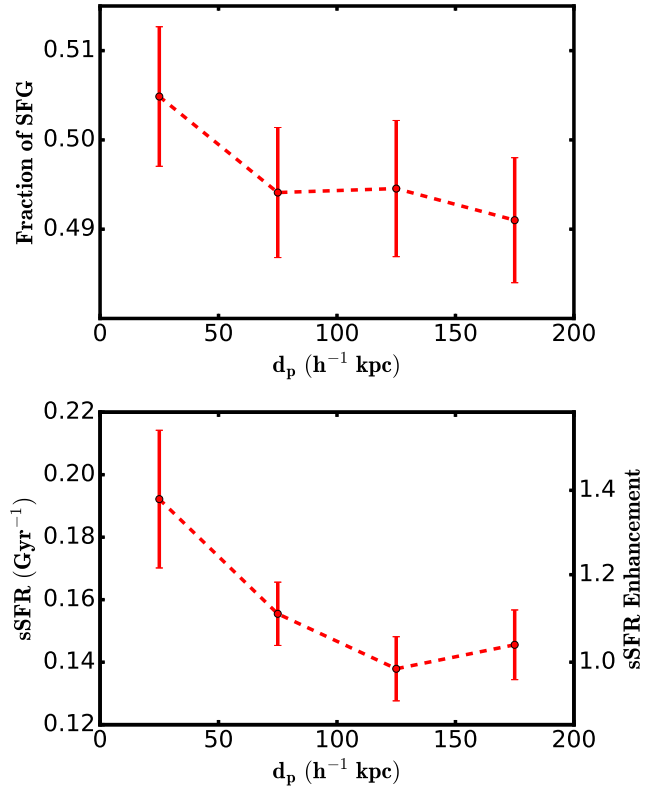}
\caption{\textit{Top panel: }the fraction of star-forming galaxy as a function of $\pdis$ for paired galaxies. \textit{Bottom panel: } the mean sSFR as a function of $\pdis$, see the text in Section \ref{sec_starformation} for details.}
\label{fig_sfr}
\end{figure}

On the other hand, the SFR enhancement of close pairs could be directly estimated from data. To do that, we take the total SFR and stellar mass of the galaxies in our galaxy pair sample from GALEX-SDSS-WISE Legacy Catalog 2 (GSWLC-2) \citep{salim2016,salim2018}. We define the galaxies which satisfy $\text{SFR}_0 / M_\star > 10^{-11} \textrm{yr}^{-1}$ as star-forming galaxies \citep{salim2018}. We show the fraction of star-forming galaxies and their mean specific SFR $(\text{sSFR}=\text{SFR}/M_\star)$ as a function of $\pdis$ for four subsamples of galaxy pairs in Figure \ref{fig_sfr}. As shown in that figure, both the fraction of star-forming galaxies and their average sSFR are roughly constant for paired galaxies when $\pdis \geq 50\kpc$. While for close pairs, both of them show increments. The increase of the fraction of the star-forming galaxies is very weak($\sim2\%$), while the enhancement of the sSFR is very significant($\sim 40\%$). Such a result confirms the scenario that the star formation efficiency of the gas-rich galaxies is significantly enriched when galaxies are in close-pair phase. The relative enhancement of the SFR we get ($\sim 40\%$) is also consistent with many earlier studies\citep{ellison2008,scudder2012,patton2013,Davies2015}. 

For a galaxy with `original' stellar mass of $M_\star$\footnote{Here, `original' means that if there is no enhanced star formation, i.e. the case that the galaxy is not in the galaxy pair environment.} , the magnitude in $r$-band brightened by the enhanced star formation could be parametrized as follows,
\begin{equation}
\Delta M_r=-2.5\log_{10}\frac{L_{r,\text{new}}+L_{r,0}}{L_{r,0}}=-2.5\log_{10}(1+\frac{\Delta M_\star/\Upsilon_{r,n}}{M_\star/\Upsilon_{r,0}}) \,
\end{equation}
where $L_{r,0}$ is the `original' luminosity of the galaxy, $\Delta M_\star$ and $L_{r,\text{new}}$ are the stellar mass and luminosity of the newborn stars, whereas $\Upsilon_{r,0}$ and $\Upsilon_{r,n}$ are the $r$-band stellar mass-to-light ratio of the `original' stellar populations and  newborn stars  respectively. The stellar mass of the newborn stars could be further parametrized by the product of the average enhanced star formation rate $\Delta \text{SFR}$ and its duration time $\tau$,  i.e. $\Delta M_\star=\Delta \text{SFR}*\tau$. However, since the amount of star formation induced by the galaxy pair environment involves many complex factors(e.g. stellar mass, gas fraction, orbit etc.), it is non-trivial to determine either $\Delta \text{SFR}$ or $\tau$ for individual galaxies in pairs.

From a statistical point of view, we assume that $\Delta \text{SFR}$ is caused by the increased star formation efficiency and parametrize it with a SFR enhancement strength $f_{\Delta}$,
\begin{equation}
f_\Delta=\frac{\Delta \text{SFR}}{\text{SFR}_0}=\frac{\Delta \text{sSFR}}{\text{sSFR}_0}\,
\end{equation}
where $\text{SFR}_0$ and $\text{sSFR}_0$ are the `original' SFR and sSFR of galaxies respectively. With this parametrization, only the star-forming(gas-rich) galaxies would show enhanced star formation, while the passive galaxies(gas-poor, $\text{SFR}_0$=0) would not. Thus, we have
\begin{equation}\label{DeltaMr}
    \begin{split}
        \Delta M_{r,L}&=-2.5\log_{10} (1+f_\Delta \tau * \text{sSFR}_0 \frac{\Upsilon_{r,0}}{\Upsilon_{r,n}})\\
        \Delta M_{r,E}&=0
    \end{split}
\end{equation}
for late(star-forming) and early(passive) type galaxies respectively.  It is well known that the sSFR of local star-forming galaxies only show a weak dependence on their stellar mass. For simplicity, we take $\text{sSFR}_0 = 1.45 \times 10^{-10} \textrm{yr}^{-1}$, which is the mean value of the paired galaxies within the bin $150 \leq \pdis \leq 200 \kpc$. 

According to Figure \ref{fig_sfr}, we take the fact that the enhanced star formation only happens in star-forming galaxies in close-pair phase. The fraction of star-forming galaxies is about 50\% and the enhancement strength $f_\Delta$ equals to 0.4. We estimate the mass-to-light ratio($\Upsilon_{r,n}$) of the newborn stars using the stellar population synthesis model of \citet{BC03}, where a continuous and constant star formation process and the Salpeter initial mass function are adopted. We calculate $\Upsilon_{r,n}$ as a function of duration time $\tau$, e.g., $\Upsilon_{r,n}=0.19$ for $\tau=0.2$ Gyr. For $\Upsilon_{r,0}$ of the `original' galaxies,  we take the approximation from \citet{blanton2003b}, where a typical color $g-r \sim 0.4$ of star-forming galaxy corresponds to a mass-to-light ratio $\Upsilon_{r,0}\sim 1.35$ \citep{bell03}.

\begin{figure}
\includegraphics[width=90mm]{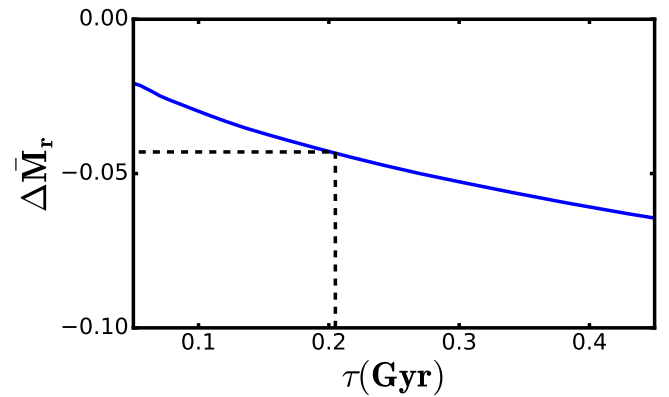}
\caption{\textit{Blue solid line:} the average brightening magnitude of pair members as a function of $\tau$ calculated from Equation \ref{DeltaMr}. \textit{Black dashed line: } the lower limit of the average brightening of $r-$band magnitude of the close pairs and its corresponding $\tau$ . }
\label{fig_dmfit}
\end{figure}

With all the above prescriptions, we can estimate the duration time $\tau$ of the SFR enhancement from the average brightening magnitude $\bar{\Delta M_r}$ of pair members in close pairs using  Equation \ref{DeltaMr}. We show $\bar{\Delta M_r}$ as a function of $\tau$ as the solid line in Figure \ref{fig_dmfit}. As can be seen, for $\bar{\Delta M_r} \geq 0.04 \text{mag}$,  the duration of the enhanced star formation is as long as $0.2$ Gyr.

The duration time $\tau \sim 0.2$ Gyr we have shown above is the average time that the close pairs have already experienced for the enhanced star formation phase. Statistically, the time-scale of the SFR enhancement phase would be two twice more, i.e., $0.4$ Gyr. Considering that the brightening magnitude we use is quite conservative, such a result seems consistent within uncertainties with numerical simulation, where the pair phase with significantly enhanced star formation lasts for about $0.1-0.5$ Gyr \citep{cox2008}. 

\subsection{Discussion}\label{sec_discussion}

In Section \ref{sec_merge_time_scale} and \ref{sec_starformation}, we have shown that the global trend of the BLFs presented in Section \ref{sec_BLFquantity} could be quantitatively explained by two different physical mechanisms, the merging time-scale effect(Section \ref{sec_merge_time_scale}) and the enhanced star formation in close-pair phase(Section \ref{sec_starformation}). However, some details of the model need to be further examined.

For the merging effect of galaxy pairs, first, we emphasize, that we adopt the relative values of the merging time-scales that have been inferred from our modeling, instead of the absolute values. Second, in a physical galaxy pair, it is the host halo mass rather than the galaxy luminosity that is the dominating factor of the merging time-scale, although the galaxy luminosity is a good proxy of its host halo mass \citep{tinker2005,mandelbaum2006,velander2013}.  Besides, the dark matter halo mass-to-light ratio of galaxies is not a constant but a non-linear relation of galaxy luminosity \citep{mandelbaum2006}. Therefore, although both the analytical formula and the parameter values of the merging time-scale used in our modeling are inferred from the simulation results, we may not make a one-to-one comparison with numerical simulations. Third, in our modeling, we have assumed a linear distribution of $\pdis$ for all configurations of galaxy pairs, which is too simplistic.  

More importantly,  our merging time-scale model is only built with a statistical approach, where only the relative number densities of galaxy pairs have been addressed.  Apparently,  our model is based on the trends of observed BLFs. We introduce the merging time-scale effect to explain the variation of BLFs, which has not dealt with any physical process of galaxy pairs. However, more realistic modeling of the $\pdis$ distribution involves a variety of complex factors, e.g., the birth rate of the galaxy pairs in the hierarchical structure, and the specific orbital parameters of pair members, which are beyond the scope of this study.  Although our model is very simple and has many uncertainties, it\textit{ indeed can distinguish }different merging time-scale models. None of the models, in which the merging time-scale only dependent on the mass ratio of pair members, could reproduce the trends of observed BLFs. To our knowledge, our simple modeling \textit{ might be} the first time that the merging time-scale of galaxy pairs could be constrained by observational data.

Switches our attention to the brightening of galaxies from enhanced star formation modeled in Section \ref{sec_starformation}, the parameterization of $\Delta M_r$ with Equation \ref{DeltaMr} is also quite simplified. First, during the parameterization of the `original' star formation activity of galaxies, we have taken the typical sSFR value of the widest galaxy pairs ($1.45 \times 10^{-10} \textrm{yr}^{-1}$)  for all star-forming galaxies. Nevertheless, the $\text{sSFR}$ of the star-forming galaxies in the local universe is not constant.  A fitting of the main sequence of SDSS galaxies shows that $\log(\text{SFR}/M_\star)=-0.35(\log M_\star-10)-9.83$ \citep{Salim2007}. Therefore, a simple constant $\text{sSFR}$ assumption might under/over estimate the brightening magnitude for low/high mass galaxies(see Equation \ref{DeltaMr}). However, as shown by Equation \ref{DeltaMr}, the effect on $\Delta M_r$ from the variation of sSFR could be partly compensated by the effect of its byproducts on $\Upsilon_0$. The galaxies with larger $\text{sSFR}$ values also have averagely bluer colors and lower mass-to-light ratios. Combining these two effects,  we have estimated that a factor of two times higher or lower $\text{sSFR}$ brings a difference of brightening magnitude  $\sim 0.02$ mag. Second, we have used only two global parameters, SFR enhancement strength $f_{\Delta}$ and duration time $\tau$, to figure out the brightening process and omit their variation among different configurations of galaxy pairs. Many observations have shown that the SFR is more significantly enhanced in major merger pairs than minor merger pairs \citep{ellison2008,Davies2015}. However, for the minor merger pairs, despite lower $f_\Delta$, they also would have larger $\tau$ values \citep{cox2008,Davies2016}. Combing these effects together, the average brightening magnitude of galaxies in pairs with different configurations may still be well approximated as a constant. Finally, it is worth mentioning that we have not taken the enhanced emissions from $\textrm{H}~\textsc{ii}$ regions and nuclear activities into account. For the newborn stars, the $\textrm{H}~\textsc{ii}$ regions only contribute  $\sim 4\%$ of $r$-band flux (through H\textsc{$\alpha$} and $[\textrm{N}~\textsc{ii}]$ emission lines, \citealt{marmol2016}) . For the nuclear activities, the fraction of optical AGNs is shown to be dependent on the galaxy pair environment very weakly \citep{argudo2016}. Despite all the uncertainties listed above and other subtle effects which we have not discussed(e.g., the fraction of star-forming galaxies in pairs), our statistical approach, which takes characteristic values of $f_\Delta$ and $\tau$ to represent the overall/average brightening of all galaxy pairs, is still quite instructive.

In summary,  the two toy models presented in this section have successfully characterized the general variation of the BLF of galaxy pairs, which bring new ideas on the study of both the galaxy merging time-scale and the star formation enhancement during the close-pair phase. To further accomplish this study, a full and detailed picture of the galaxy pair evolution is required, which could be gradually manifested through the combinations of the large sample statistics(e.g. this study), the state-of-art numerical simulations(e.g. cosmological simulation with baryon physics, \citealt{rg2015,sparre2016}), and the detailed observations of the galaxy pairs in different merging phases (e.g. integral-field-spectroscopy study, \citealt{yuan2018,fu2018}).

\section{summary}\label{sec_sum}

In this study, we have compiled a sample of $46,510$ isolated galaxy pairs based on the SDSS Main Galaxy Sample. We also benefit from a significant number of extra redshifts from other surveys(e.g., the LAMOST spectral and GAMA surveys). The completeness of $\theta \leq 55{''}$ pairs in our sample is significantly improved compared to previous SDSS-based galaxy pair samples. 

Based on our large galaxy pair sample, we have calculated the BLF of galaxy pairs and studied its dependence on the projected distance between pair members $\pdis$ in detail. We find that the BLF degenerates into the LF of single galaxies at large distance ($\pdis \geq 150\kpc$), which indicates that galaxy-galaxy interaction starts from $\pdis \leq 150\kpc$. At $50\kpc \leq \pdis \leq 150\kpc$, the BLFs deviate from the LFs of single galaxies significantly, where both the total magnitude of galaxy pairs and magnitude gap between pair members increase as $\pdis$ decrease. By integrating the BLFs of galaxy pairs, we also have obtained the ULF of galaxy pairs. Comparing with BLFs, the ULFs only show weak dependence on $\pdis$.

The BLF variation is consistent with a physical scenario where the dynamic friction leads to massive and major merger pairs more rapidly merging. For the closest galaxy pairs ($10 \kpc \leq \pdis \leq 50 \kpc$), the average magnitude of pair members shows an overall brightening trend, which manifests the enhanced star formation in this close-pair phase. It might be the first time that our statistical study reveals observational evidence for the merging time-scale of galaxy pairs. Our conclusion on the merging time-scale approximation, $T_{\textrm{m}} \propto (L_1/L_2)^{0.4} (L_1+L_2)^{-0.33}$, qualitatively agrees with the conclusion from dark matter only numerical simulations of \citet{Jiang2014}. From the overall brightening of $\Delta M_r \approx 0.04$ mag for the close pairs, we conclude that the starburst time-scale is about $0.4$ Gyr considering $40\%$ sSFR enhancement, which strengthens the conclusions from other observational studies(e.g. \citealt{woods2010}) and numerical simulations(e.g. \citealt{cox2008}). 

In this study, we have only calculated the BLFs of galaxy pairs in the SDSS $r$-band. In principle, such measurements can be applied to other SDSS bands. Combining the BLFs in multi-bands will provide further constraints and insights on the galaxy-galaxy interaction in galaxy pairs. With the on-going data release of LAMOST spectral survey, the completeness of the galaxy pair sample can be further improved. Then, we expect a future study of the BLFs of galaxy pairs in other SDSS bands and for other detailed sub-samples(e.g., different morphology configurations).

\acknowledgments

This work is supported by the National Natural Science Foundation of China (No. 11573050 and 11433003) and the Science Foundation of Shanghai (No. 16ZR1442100). The authors thank Hugh Jones for carefully reading the manuscript and useful suggestions.

%This work has made use of the \texttt{PYTHON} packages \texttt{NUMPY} \citep{numpy}, \texttt{SCIPY} \citep{scipy}, \texttt{MATPLOTLIB} \citep{matplotlib}, \texttt{emcee} \citep{emcee} and \texttt{ASTROPY} \citep{astropy}.

This work has made use of data products from the Sloan Digital Sky Survey (SDSS, \url{http://www.sdss.org}), the Large Sky Area Multi-Object Fibre Spectroscopic Telescope (LAMOST, \url{http://www.lamost.org}), the GAMA survey(\url{http://www.gama-survey.org}). Thanks for their tremendous efforts on the surveying work.

Guoshoujing Telescope (the Large Sky Area Multi-Object Fiber Spectroscopic Telescope LAMOST) is a National Major Scientific Project built by the Chinese Academy of Sciences. Funding for the project has been provided by the National Development and Reform Commission. LAMOST is operated and managed by the National Astronomical Observatories, Chinese Academy of Sciences.

\bibliographystyle{aasjournal}
\bibliography{ref.bib}

\end{CJK*}
\end{document}